\documentclass[10pt]{article}

\usepackage[english]{babel}

\usepackage[pdftex]{graphics,color}
\usepackage{epsfig}

\begin{document}
\date{}
\begin{center}
{\Large\bf A simple model for a minimal environment: the two-atom Tavis-Cummings model revisited
}
\end{center}
\begin{center}
{\normalsize G.L. De\c cordi and A. Vidiella-Barranco \footnote{vidiella@ifi.unicamp.br}}
\end{center}
\begin{center}
{\normalsize{ Instituto de F\'\i sica ``Gleb Wataghin'' - Universidade Estadual de Campinas}}\\
{\normalsize{ 13083-859   Campinas  SP  Brazil}}\\
\end{center}
\begin{abstract}

Individual quantum systems may be interacting with surrounding environments having a small 
number of degrees of freedom. It is therefore relevant to understand the extent to which such 
small (but uncontrollable) environments could affect the quantum properties of the system of 
interest. Here we discuss a simple system-environment toy model, constituted by a two-level atom 
(atom 1) interacting with a single mode cavity field. The field is also assumed to be (weakly) 
coupled to an external noisy subsystem, the small environment, modeled as a second two-level atom (atom 2). 
We investigate the action of the minimal environment on the dynamics of the linear entropy (state purity) 
and the atomic dipole squeezing of atom 1, as well as the entanglement between atom 1 and the field.
We also obtain the full analytical solution of the two atom Tavis-Cummings model for both arbitrary coupling 
strengths and frequency detunings, necessary to analyze the influence of the field-environment detuning on 
the evolution of the above mentioned quantum properties. For complementarity, we discuss the role of the degree 
of mixedness of the environment by analyzing the time-averaged linear entropy of atom 1.
 
\end{abstract}
\section{Introduction}

The evolution of quantum systems in contact with external environments has been the subject of investigation for quite some time 
\cite{louisell73}. The environment is usually modeled as a (quantum) system having a large number of degrees of 
freedom, Viz.,  a reservoir, and number of methods, mainly perturbative, have been developed in order to study its influence 
on the behaviour of quantum systems \cite{louisell73}. It is in general possible to derive an equation describing the 
evolution of the reduced density operator of the system of interest - the master equation, obtained by tracing over the degrees of 
freedom of the environment. Generally speaking, the coupling to an external environment has a detrimental action on the quantum 
properties of the system of interest, causing effects such as the loss of quantum coherence (decoherence) \cite{cald85,milb85,zeh96}. 
The models of environment mostly rely on the assumption of the existence of an ideal (large) reservoir, which naturally leads to 
irreversibility and relaxation features. For a system of interest being a single mode (cavity) field of the quantized field, 
cavity losses (dissipation) may be modeled via ideal reservoirs constituted by a large collection of either independent field 
modes \cite{louisell73} or a beam of two-level atoms \cite{lambscully74}. Both procedures lead basically to the same master equation 
for the reduced density operator of the cavity field. 

With the recent advances of quantum technologies \cite{review17a,review17b}, the investigation of the influence of 
the surrounding environments on the evolution of quantum systems became particularly important. 
Even if the environment has a small number of degrees of freedom, 
one expects that those unwanted couplings will affect in some way the quantum properties of the 
system of interest, as well as bring memory effects. Remarkably, a ``minimal environment" constituted by a single 
electron may strongly affect a coherent signature, i.e., the interference fringes related to a second electron 
(system of interest), as has been experimentally demonstrated in the double photoionization of $\mbox{H}_2$ molecules 
[see reference \cite{akoury07}]. Besides, the effects of small environments (e.g., a single 
harmonic oscillator) have also been addressed 
in \cite{hanggi09,avb14,ashhab14,avb16}. As a matter of fact, one of the 
smallest possible environments could be constituted by a single two-level system, opposed to the model of a reservoir containing 
a large number of atoms \cite{lambscully74}. In \cite{avb14} it is studied a system made of a two-level system (atom 1) coupled to 
an oscillator (field mode) which is itself in interaction with a minimal environment constituted by a second two-level system (atom 2). 
This is the well-known two-atom Tavis-Cummings model (TCM) \cite{cumm68}, but where an asymmetric partition of the 
system has been considered; the system of interest being constituted by atom 1 + field, while a partial trace is performed over the 
environment (atom 2). In \cite{avb14}, the discussion is restricted to the exact resonance case, i.e, the atom 1(2)-field detunings 
being equal ($\Delta_1 = \Delta_2 = 0$), although the atoms are assumed to be coupled with different strengths to the field 
($\lambda_1 \neq \lambda_2$). If the field is completely isolated from atom 2, i.e., $\lambda_2 = 0.0$, we have the basic 
Jaynes-Cummings model \cite{jaynes63,knight93}, a particular case of the TCM ($N = 1$). The Jaynes-Cummings model has well known 
features; for instance, coherent Rabi oscillations of the atomic inversion as well as the linear entropy of the atom, 
$S = 1 - Tr(\rho^2_a)$, \footnote{Being $\rho_a$ the reduced atomic density operator.} if the field is initially prepared in a Fock 
state. Such a characteristic behavior may be useful to evaluate the influence of external systems on the regular evolution of a 
quantum system, and thus we could focus our attention on the reduced density operator of atom 1, $\rho_{a1}$. 
Here we investigate in which way some quantum features of the system of interest are affected due to its coupling 
to a small, but noisy environment. We analyze the evolution of non-classical effects like the atomic dipole squeezing 
\cite{zoller81,raz93}, as well the entanglement between atom 1 and the field, given that in general non-classical effects are 
susceptible to external influences.
Besides, we also study the influence of the detuning between the field and the environment on the evolution of the system.
We note that the analytical solutions of the two-atom TCM found in the literature are restricted to particular cases. Namely, 
the atom-field coupling constants may be assumed to be equal (identical atoms) \cite{deng85}, or different (non-identical atoms),
but still having the frequency of the field equal to the atomic transition frequencies  \cite{zub87,raz93}. 
A step further is given in \cite{huy94}, where it is presented a solution of the model for non-identical 
atoms and non-zero detuning. However the authors consider the same detuning for both atoms ($\Delta_1 = \Delta_2 = \Delta$). 
As we would like to explore a more general situation, we worked out an exact analytical solution of the two-atom TCM for 
distinct coupling constants ($\lambda_1,\lambda_2$) and arbitrary detunings ($\Delta_1,\Delta_2$). Such a general solution 
provides us additional flexibility, and we may treat the case in which the environment is detuned from the field while 
atom 1 remains in resonance with it. Yet, in a realistic set-up we may have some control over $\Delta_2$, which in principle 
would allow partial restoration of the quantum properties of the system. Depending on the property we are interested in,  
a different amount of detuning may be required. Of course for a very large detuning the environment would be effectively 
decoupled from the field, allowing the full restoration of every quantum property of the system.
Our paper is organized as follows. In section 2 we obtain the general analytical solution of the 
two-atom TCM with arbitrary detunings and couplings. In section 3 we discuss the time evolution of the linear entropy, 
the atomic dipole squeezing of atom 1 as well as the atom 1-field entanglement, considering atom 2 as a disturbance 
(environment). We also show how it would be possible to restore quantum coherence, the atomic dipole squeezing and entanglement 
by controlling the frequency detuning between the field and the environment (atom 2). In section 4 we make some considerations 
about the time-averaged linear entropy, and in section 5 we summarize our conclusions. Some details of the calculations are
shown in the Appendix. 

\section{Tavis-Cummings model: an analytical solution for different coupling constants and different detunings}

The two-atom TCM is described by the following Hamiltonian (under the rotating wave approximation and making $\hbar = 1$)

\begin{equation}
H=\frac{\omega_{1}}{2}\sigma_{1}^{z}+\frac{\omega_{2}}{2}\sigma_{2}^{z}+\omega a^{\dagger}a+\lambda_{1}(a\sigma_{1}^{+}+a^{\dagger}\sigma_{1}^{-})+\lambda_{2}(a\sigma_{2}^{+}+a^{\dagger}\sigma_{2}^{-})\,,\label{eq:HTC completo}
\end{equation}
where $a\left(a^{\dagger}\right)$ are the annihilation
(creation) operators associated to the field mode, with frequency $\omega$;
$\sigma_{i}^{z}\,,\sigma_{i}^{+}$ e $\sigma_{i}^{-}$ are the
de Pauli operators relative to the ``i-th" atom, each atom having transition frequency $\omega_{i}$. We may rewrite the
Hamiltonian above in terms of the detunings $\Delta_{1}=\omega_{1}-\omega$ and $\Delta_{2}=\omega_{2}-\omega$, as
\begin{equation}
H = \omega\left(\frac{\sigma_{1}^{z}+\sigma_{2}^{z}}{2}+a^{\dagger}a\right)+\frac{\Delta_{1}}{2}\sigma_{1}^{z}+\frac{\Delta_{2}}{2}\sigma_{2}^{z}+\lambda_{1}(a\sigma_{1}^{+}+a^{\dagger}\sigma_{1}^{-})+\lambda_{2}(a\sigma_{2}^{+}+a^{\dagger}\sigma_{2}^{-})\;.
\label{eq:HTC reescrito}\\ 
\end{equation}
The Hamiltonian (\ref{eq:HTC reescrito}) may be split in two parts
\begin{equation}
H=H_{0}+H_{1},
\end{equation}
where
\begin{equation}
H_{0}=\omega\left(\frac{\sigma_{1}^{z}+\sigma_{2}^{z}}{2}+a^{\dagger}a\right)
\end{equation}
is related to a conserved quantity, and
\begin{equation}
H_{1}=\frac{\Delta_{1}}{2}\sigma_{1}^{z}+\frac{\Delta_{2}}{2}\sigma_{2}^{z}+\lambda_{1}(a\sigma_{1}^{+}+a^{\dagger}\sigma_{1}^{-})
+\lambda_{2}(a\sigma_{2}^{+}+a^{\dagger}\sigma_{2}^{-})\;
\end{equation}
is the interaction part. 

The Schr{\"o}dinger equation in the interaction representation reads
\begin{equation}
i\,\frac{d}{dt}\left|\psi_{I}\left(t\right)\right\rangle =H_{I}\left|\psi_{I}\left(t\right)\right\rangle \;,\label{eq:Sc}
\end{equation}
with $H_{I}=\exp\left(i\, H_{0}t\right)H_{1}\exp\left(-i\, H_{0}t\right)$, and $H_{I}=H_{1}$. 

The Hamiltonian $H_{I}$ induces transitions between the states 
$\left|e_{1},e_{2},n\right\rangle $, $\left|e_{1},g_{2},n+1\right\rangle $,
$\left|g_{1},e_{2},n+1\right\rangle $, $\left|g_{1},g_{2},n+2\right\rangle $,
and therefore we may write the following {\it ansatz}
\begin{eqnarray}
\left|\psi_{I}\left(t\right)\right\rangle &=&C_{1,n}\left(t\right)\left|e_{1},e_{2},n\right\rangle 
+C_{2,n}\left(t\right)\left|e_{1},g_{2},n+1\right\rangle \nonumber\\ 
&+&C_{3,n}\left(t\right)\left|g_{1},e_{2},n+1\right\rangle 
+C_{4,n}\left(t\right)\left|g_{1},g_{2},n+2\right\rangle \;.\label{eq:astz}
\end{eqnarray}
After substituting the proposed solution (\ref{eq:astz}) above in the Schr{\"o}dinger equation, we obtain the corresponding
set of coupled differential equations for the amplitudes $C_{i,n}(t)$
\begin{eqnarray}
i\,{\dot C}_{1,n} & = & \left(\frac{\Delta_{1}+\Delta_{2}}{2}\right)C_{1,n}+\lambda_{2}\sqrt{n+1}\, C_{2,n}
+\lambda_{1}\sqrt{n+1}\, C_{3,n}\,,\nonumber \\
i\,{\dot C}_{2,n} & = & \lambda_{2}\sqrt{n+1}\, C_{1,n}+\left(\frac{\Delta_{1}-\Delta_{2}}{2}\right)C_{2,n}
+\lambda_{1}\sqrt{n+2}\, C_{4,n}\,,\nonumber \\
i\,{\dot C}_{3,n} & = & \lambda_{1}\sqrt{n+1}\, C_{1,n}+\left(\frac{\Delta_{2}-\Delta_{1}}{2}\right)C_{3,n}
+\lambda_{2}\sqrt{n+2}\, C_{4,n}\,,\nonumber \\
i\,{\dot C}_{4,n} & = & \lambda_{1}\sqrt{n+2}\, C_{2,n}+\lambda_{2}\sqrt{n+2}\, C_{3,n}-\left(\frac{\Delta_{1}
+\Delta_{2}}{2}\right)C_{4,n}\,. \label{eq:acopladas}
\end{eqnarray} 

In order to solve the system of coupled differential equations (\ref{eq:acopladas}), we have employed the Laplace 
transform method, and the set of differential equations is transformed to the following set of algebraic equations
\begin{eqnarray}
\left(-\frac{\Delta_{1}+\Delta_{2}}{2}+i\, s\right)\tilde{C_{1}}-\lambda_{2}\sqrt{n+1}\,\tilde{C}_{2}-\lambda_{1}\sqrt{n+1}\,\tilde{C}_{3} & = & i\, c_{1}\left(0\right)\nonumber \\
-\lambda_{1}\sqrt{n+1}\,\tilde{C}_{1}+\left(\frac{\Delta_{2}-\Delta_{1}}{2}+i\, s\right)\tilde{C}_{2}-\lambda_{1}\sqrt{n+2}\,\tilde{C}_{4} & = & i\, c_{2}\left(0\right)\nonumber \\
-\lambda_{1}\sqrt{n+1}\,\tilde{C}_{1}+\left(\frac{\Delta_{1}-\Delta_{2}}{2}+i\, s\right)\tilde{C}_{3}-\lambda_{2}\sqrt{n+2}\,\tilde{C}_{4} & = & i\, c_{3}\left(0\right)\nonumber \\
-\lambda_{1}\sqrt{n+2}\,\tilde{C}_{2}-\lambda_{2}\sqrt{n+2}\,\tilde{C}_{3}+\left(\frac{\Delta_{1}+\Delta_{2}}{2}+i\, s\right)\tilde{C}_{4} & = & i\, c_{4}\left(0\right).\nonumber
\end{eqnarray}

For simplicity we have denoted $\tilde{C_{j}}\left(s\right)=\tilde{C_{j}}$ in the equations above. 
The subsequent steps involve the solution of polynomials up to fourth degree, and after some involved calculations 
we obtain the full solution. Curiously this leads to much more intricate formulae compared to the equal detunings case 
($\Delta_1 = \Delta_2$), and as the expressions are rather lengthy, we have included them in the Appendix. 

\section{Numerical results: quantum state purity, dipole squeezing and entanglement}

We discuss now the reduced dynamics of the quantum system, basically focusing on the properties associated to atom 1. 
In order to do so, we should first trace the total density operator $\rho(t)$ over the variables of atom 2 (the ``environment"), 
obtaining the joint atom 1-field density operator, or $\rho(t)_{a1,f} = Tr_{a2}\rho(t)$. For an isolated 
atom (Jaynes Cummings model, \cite{jaynes63}), we know that the linear entropy has a completely reversible behaviour, for the field
initiallly prepared in a Fock state. Moreover, non-classical features such as squeezing \cite{zoller81} may also arise during the 
atom-field interaction. If we want to focus solely on the atomic properties, we should perform a further partial trace over the field 
variables, i.e., calculate the reduced density operator relative to atom 1,
$\rho(t)_{a1} = Tr_{f}\left[\rho(t)_{a1,f}\right]$.
Here we are going to assume initial conditions of the form:
$\rho\left(0\right)=\rho_{a1}\left(0\right)\otimes\rho_{f}\left(0\right)\otimes\rho_{a2}\left(0\right)$, with
\begin{equation}
\rho_{a1}\left(0\right)=\left|\phi_{1}\right\rangle \left\langle \phi_{1}\right|\,,
\quad\rho_{f}\left(0\right)=\left|N\right\rangle \left\langle N\right|\,,
\quad\rho_{a2}\left(0\right)=p\left|e_{2}\right\rangle \left\langle e_{2}\right|
+\left(1-p\right)\left|g_{2}\right\rangle \left\langle g_{2}\right|\,.\label{eq:est inicial}
\end{equation}
In other words, having atom 1 prepared in the state $\left|\phi_{1}\right\rangle =\cos\left(\frac{\theta}{2}\right)\left|g_{1}\right\rangle 
+\sin\left(\frac{\theta}{2}\right)e^{i\phi}\left|e_{1}\right\rangle $ and the field prepared in 
a Fock state $\left|N\right\rangle$, with $N$ being either $N=1$ or $N=0$.
In this section we are going to address the case of a maximally mixed environment, i.e., atom 2 initially in a 
statistical mixture of its ground and excited states with $p = 1/2$, or
$\rho_{a2}(0)=\frac{1}{2}\left(\left|g_{2}\right\rangle \left\langle g_{2}\right|+\left|e_{2}\right\rangle \left\langle e_{2}\right|\right)$.
We may now discuss some of the quantum dynamical features of the system. 

\subsection{Reduced dynamics of atom 1: linear entropy evolution}

The linear entropy is used to quantify the degree of purity of a quantum state. In the case of atom 1, it may be written as
\begin{equation}
S\left(t\right)=1-Tr_{a1}\left(\rho_{a1}^{2}\right)=2\left[\alpha-\alpha^{2}-\left|\gamma\right|^{2}\right]\,,\label{eq:entropiatc}
\end{equation}
where
\begin{eqnarray}
\alpha\left(t\right)  =  \sin^{2}\left(\frac{\theta}{2}\right)\left\{ p\left[\left|C_{3,1}^{\left(I\right)}\right|^{2}+
\left|C_{4,1}^{\left(I\right)}\right|^{2}\right]+\left(1-p\right)\left[\left|C_{3,0}^{\left(II\right)}\right|^{2}+
\left|C_{4,0}^{\left(II\right)}\right|^{2}\right]\right\} \nonumber \\
\label{eq:pop ground}\\
+\cos^{2}\left(\frac{\theta}{2}\right)\left\{ p\left[\left|C_{3,0}^{\left(III\right)}\right|^{2}+
\left|C_{4,0}^{\left(III\right)}\right|^{2}\right]+\left(1-p\right)\left[\left|C_{3,-1}^{\left(IV\right)}\right|^{2}+
\left|C_{4,-1}^{\left(IV\right)}\right|^{2}\right]\right\}, \nonumber 
\end{eqnarray}
and
\begin{eqnarray}
\gamma\left(t\right)  =  e^{-i\phi}\frac{\sin\left(\theta\right)}{2}\left\{ 
p\left[C_{1,1}^{\left(I\right)^{*}}C_{3,0}^{\left(III\right)}
+C_{2,1}^{\left(I\right)^{*}}C_{4,0}^{\left(III\right)}\right]+\right.\nonumber \\
\label{eq:coherence tc}\\
\left.+\left(1-p\right)\left[C_{1,0}^{\left(II\right)^{*}}C_{3,-1}^{\left(IV\right)}
+C_{2,0}^{\left(II\right)^{*}}C_{4,-1}^{\left(IV\right)}\right]\right\} \nonumber 
\end{eqnarray}
are the populations and coherences (respectively) of the reduced density operator of atom 1
\begin{equation}
\rho_{a1}\left(t\right)=\alpha\left|g_{1}\right\rangle \left\langle g_{1}\right|
+(1-\alpha)\left|e_{1}\right\rangle \left\langle e_{1}\right|+\gamma\left|g_{1}\right\rangle \left\langle e_{1}\right|
+\gamma^{*}\left|e_{1}\right\rangle \left\langle g_{1}\right|\,.\label{eq:ro reduzido}
\end{equation}
Here the superscript $i$, $\left(i=I,\, II,\, III,\, IV\right)$ in the amplitudes $C_{j,n}^{\left(i\right)}$ in equations 
(\ref{eq:pop ground}) and (\ref{eq:coherence tc}) above are related to the initial conditions. The amplitudes
$C_{j,n}\left(t\right)$ may be written as a linear combination
of the initial conditions as $C_{j,n}\left(t\right)=\sum_{j,\, m=1}^{4}A_{j\, m}c_{j}\left(0\right)$.
The notation employed is such that $C_{1,1}^{\left(I\right)}$, for instance, indicates that
$C_{1}\left(0\right)=1$ and the remaining coefficients are zero; the superscript $\left(II\right)$ in $C_{2,0}^{\left(II\right)}$ 
indicates that $C_{2}\left(0\right)=1,$ and the remaining coefficients are zero and so on.

\begin{figure}
\begin{centering}
\includegraphics[scale=0.9]{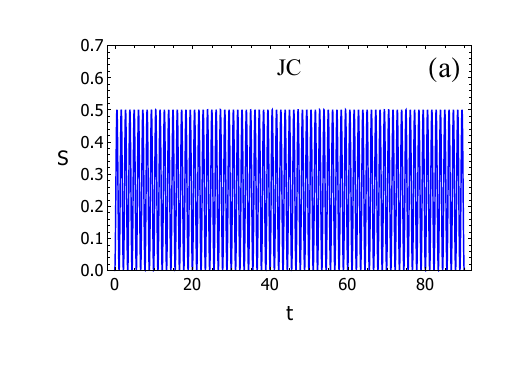}\includegraphics[scale=0.9]{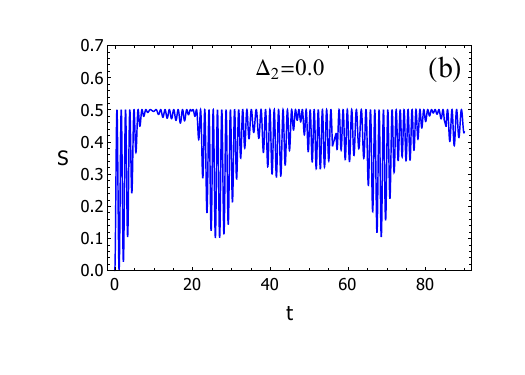}
\includegraphics[scale=0.9]{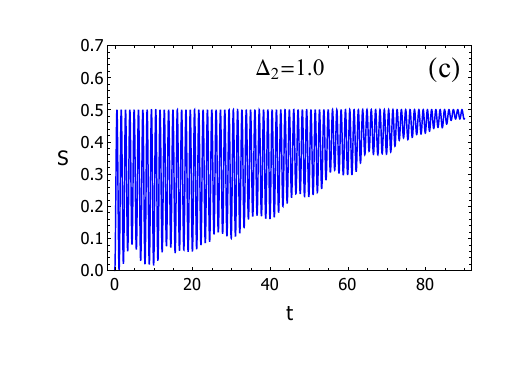}\includegraphics[scale=0.9]{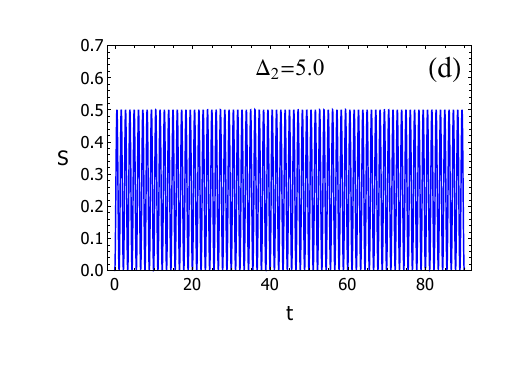}
\par\end{centering}

\caption{\label{fig:tc entropia}Linear entropy of atom 1 as a function of
time. In (a) the field is not coupled to the environment, i.e.,
$\lambda_{2}=0.0$ (Jaynes Cummings model). For the field coupled to the environment, with $\lambda_{2}=0.1$
and (b) $\Delta_{2}=0.0$; (c) $\Delta_{2}=1.0$, and (d) $\Delta_{2}=5.0$.
In all cases $\lambda_{1}=1.0$, and the initial state of the system is a tripartite product state with
$\left|\psi_{a1}\right\rangle =\left|e_{1}\right\rangle $,
$\rho_{a2}=\frac{1}{2}\left(\left|g_{2}\right\rangle \left\langle g_{2}\right|+\left|e_{2}\right\rangle \left\langle e_{2}\right|\right)$
and $\left|\psi_{f}\right\rangle =\left|1\right\rangle $. }\label{fig1}
\end{figure}

We would like now to analyze the effect of $\Delta_2$ (atom 2-field detuning) on the time evolution of the linear entropy 
of atom 1. In figure (\ref{fig1}) we have plots of the linear entropy of atom 1 as a function of time considering different values of
$\Delta_2$. The field is initially prepared in the one photon Fock state $\left|1\right\rangle$, atom 1 in its excited state 
$|e_1\rangle$, and atom 2 in a maximally mixed state, 
$\rho_{a2}(0)=\frac{1}{2}\left(\left|g_{2}\right\rangle \left\langle g_{2}\right|+\left|e_{2}\right\rangle \left\langle e_{2}\right|\right)$.

As seen in figure (\ref{fig1}a), the linear entropy is a periodic function of time (Rabi oscillations) if atom 2 is decoupled from the field, 
i.e., for $\lambda_2 = 0.0$. If the interaction between atom 2 and the field is turned on ($\lambda_2 = 0.1$), the evolution of the linear
entropy becomes very irregular, as shown in figure (\ref{fig1}b), which characterizes a destructive effect due to an unwanted coupling
to a noisy sub-system. Nevertheless, it is possible to restore periodicity by controlling the atom 2-field frequency detuning. 
If $\Delta_2$ is increased, we expect less influence of atom 2 over the dynamics of the system of interest. In fact, as shown in figure 
(\ref{fig1}c) and figure (\ref{fig1}d), periodicity may be re-established for a sufficiently large $\Delta_2$. We remark that atom 1
is kept on resonance with the field ($\Delta_1 = 0.0$) as well as strongly coupled to it ($\lambda_1 = 1.0$).

\subsection{Reduced dynamics of atom 1: atomic dipole squeezing}

Atomic dipole squeezing is a non-classical effect that may arise in the Jaynes-Cummings model \cite{zoller81}. 
An atom is said to be dipole squeezed if the quantum fluctuations of the atomic dipole are below the fundamental limit imposed 
by the Heisenberg inequalities. The components of the (slowly varying) atomic dipole operator (for atom 1) may be written 
as \cite{zoller81}

\begin{eqnarray}
\sigma_{x} & = & \sigma_{1}^{+}e^{i\omega_{1}t}+\sigma_{1}^{-}e^{-i\omega_{1}t}\,,\nonumber \\
\sigma_{y} & = & \frac{1}{i}\left(\sigma_{1}^{+}e^{i\omega_{1}t}-\sigma_{1}^{-}e^{-i\omega_{1}t}\right)\,.
\end{eqnarray}
The operators above do not commute ($\left[\sigma_{x},\sigma_{y}\right]=2i\,\sigma_{z}$), 
and thus they should obey the Heisenberg inequality 
$\left(\Delta\sigma_{x}\right)\left(\Delta\sigma_{y}\right)\geq\left|\left\langle \sigma_{z}\right\rangle \right|$.
Atomic dipole squeezing is verified if
\begin{equation}
\left(\Delta\sigma_{x,\, y}\right)^{2}<\left|\left\langle \sigma_{z}\right\rangle \right|\,,
\end{equation}
where $\left(\Delta\sigma_{x}\right)^{2}=1-4\left(Re\left\langle \sigma_{1}^{-}\right\rangle e^{-i\,\omega_{1}t}\right)^{2}$
and $\left(\Delta\sigma_{y}\right)^{2}=1-4\left(Im\left\langle \sigma_{1}^{-}\right\rangle e^{-i\,\omega_{1}t}\right)^{2}$.
The conditions for squeezing may be written in terms of the functions $s_{1}$ e $s_{2}$ (indexes of squeezing) as
\begin{equation}
s_{1}=\frac{1-4\left(Re\left\langle \sigma_{1}^{-}\right\rangle 
e^{-i\,\omega_{1}t}\right)^{2}}{\left|\left\langle \sigma_{z}\right\rangle \right|}<1\quad \mbox{or}\quad 
s_{2}=\frac{1-4\left(Im\left\langle \sigma_{1}^{-}\right\rangle 
e^{-i\,\omega_{1}t}\right)^{2}}{\left|\left\langle \sigma_{z}\right\rangle \right|}<1\;,\label{eq:def squeezing}
\end{equation}
where $\omega_{1}$ is the frequency of atom 1, and $\left\langle \sigma_{z}\right\rangle $
is its corresponding atomic inversion. Here $\left\langle \sigma_\alpha\right\rangle =Tr_{1}\left(\sigma_\alpha\rho_{a1}\right)$.

We may rewrite the indexes of squeezing, $s_1$ and $s_2$, in terms of the functions $\alpha(t)$ and $\gamma(t)$ above, or 
\begin{eqnarray}
s_{1}&=&\frac{1-4\left(Re[\gamma]\cos\left(\omega_{1}t\right)-Im[\gamma]\sin\left(\omega_{1}t\right)\right)^{2}}{\left|1-2\alpha\left(t\right)\right|}<1\quad,\nonumber
\\
s_{2}&=&\frac{1-4\left(Re[\gamma]\sin\left(\omega_{1}t\right)+Im[\gamma]\cos\left(\omega_{1}t\right)\right)^{2}}{\left|1-2\alpha\left(t\right)\right|}<1\;.
\label{indexes}
\end{eqnarray}

\begin{figure}
\begin{centering}
\includegraphics[scale=0.9]{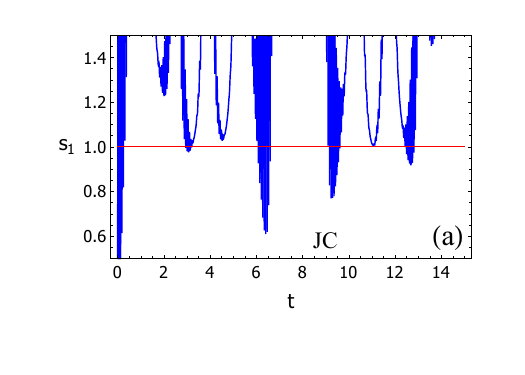}\includegraphics[scale=0.9]{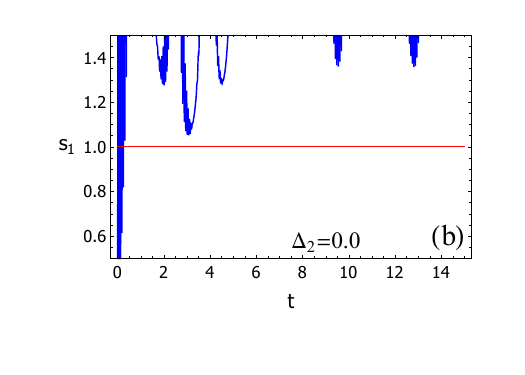}
\includegraphics[scale=0.9]{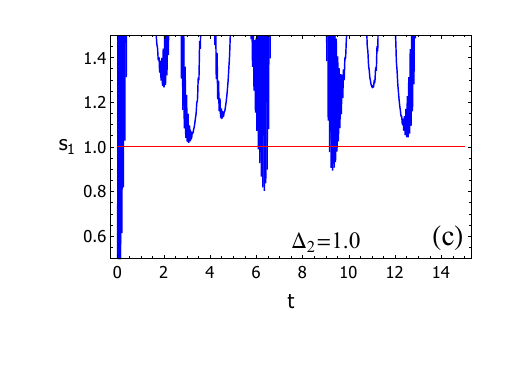}\includegraphics[scale=0.9]{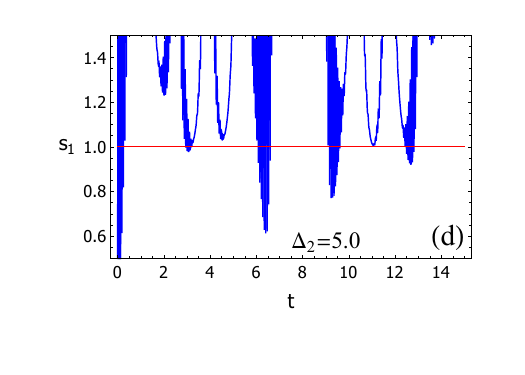}
\par\end{centering}

\caption{\label{fig:tc squeezing}Squeezing index $s_{1}$ as a function of
time. In (a) the field is not coupled to the environment, i.e.,
$\lambda_{2}=0.0$ (Jaynes Cummings model). For the field coupled to the environment, with $\lambda_{2}=0.1$
and (b) $\Delta_{2}=0.0$; (c) $\Delta_{2}=1.0$, and (d) $\Delta_{2}=5.0$.
In all cases $\lambda_{1}=1.0$, and the initial state of the system is a tripartite product state with$\left|\psi_{a1}\right\rangle =\cos\left(0.6\right)\left|g_{1}\right\rangle +\sin\left(0.6\right)\left|e_{1}\right\rangle $,
$\rho_{a2}=\frac{1}{2}\left(\left|g_{2}\right\rangle \left\langle g_{2}\right|+\left|e_{2}\right\rangle \left\langle e_{2}\right|\right)$
and $\left|\psi_{f}\right\rangle =\left|1\right\rangle $. }\label{fig2}
\end{figure}

Now we choose specific initial conditions which allow dipole squeezing in the case of having atom 2 completely decoupled from the field
($\lambda_2 = 0.0$), which corresponds to the Jaynes-Cummings model. Atom 1 is assumed to be prepared in the superposition state 
$\left|\phi_{1}(0)\right\rangle =\cos\left(0.6\right)\left|g_{1}\right\rangle +\sin\left(0.6\right)\left|e_{1}\right\rangle $;
atom 2 in the maximally mixed state, and the field in the one photon Fock state, $\left|1\right\rangle $. 
In figure (\ref{fig2}) we have plotted the dipole squeezing index $s_{1}$ relative to atom 1 as a function of time. 
We note that in the absence of the ``environment" (atom 2), dipole squeezing occurs for a few narrow intervals of time 
[figure (\ref{fig2}a)]. If atom 2 is weakly coupled to the field, e.g.,
$\lambda_{2}=0.1$, the dipole squeezing is inhibited due to the action of the environment [see figure (\ref{fig2}b)].
Yet, similarly to what we have seen in the previous subsection, dipole 
squeezing may be restored for a large enough detuning between atom 2 and the field, as shown in figures (\ref{fig2}c) and (\ref{fig2}d).

\subsection{Atom 1-field entanglement}

We may quantify the entanglement between atom 1 and the field by evaluating the negativity \cite{negativity}, an
entanglement measure convenient in this case. Now we assume the following initial conditions:
the field initially prepared in the vacuum state $\left|0\right\rangle$, atom 1 in its excited state 
$|e_1\rangle$, and atom 2 in a maximally mixed state.
The negativity may be calculated from the time-dependent atom 1-field reduced density operator
\begin{eqnarray*}
\rho_{a1-f}\left(t\right) & = & \rho_{11}\left|g_{1},0\right\rangle \left\langle g_{1},0\right|+\rho_{22}\left|g_{1},1\right\rangle \left\langle g_{1},1\right|+\rho_{33}\left|g_{1},2\right\rangle \left\langle g_{1},2\right|+\\
 &  & +\rho_{44}\left|e_{1},0\right\rangle \left\langle e_{1},0\right|+\rho_{55}\left|e_{1},1\right\rangle \left\langle e_{1},1\right|
+\rho_{24}\left|g_{1},1\right\rangle \left\langle e_{1},0\right|+\\
 &  & +\rho_{24}^{*}\left|e_{1},0\right\rangle \left\langle g_{1},1\right|
+\rho_{35}\left|g_{1},2\right\rangle \left\langle e_{1},1\right|+\rho_{35}^{*}\left|e_{1},1\right\rangle \left\langle g_{1},2\right|.
\end{eqnarray*}
The relevant matrix elements are

\[
\rho_{11}=\frac{1}{2}\left|C_{3,-1}^{\left(II\right)}\right|^{2}\,,\quad\quad\rho_{22}=\frac{1}{2}\left(\left|C_{3,0}^{\left(I\right)}\right|^{2}+\left|C_{4,-1}^{\left(II\right)}\right|^{2}\right)\,,
\]

\[
\rho_{33}=\frac{1}{2}\left|C_{4,0}^{\left(I\right)}\right|^{2}\,,\qquad\rho_{44}=\frac{1}{2}\left(\left|C_{1,0}^{\left(I\right)}\right|^{2}+\left|C_{2,-1}^{\left(II\right)}\right|^{2}\right)\,,
\]

\[
\rho_{55}=\frac{1}{2}\left|C_{2,0}^{\left(I\right)}\right|^{2}\,,\qquad\rho_{35}=\frac{1}{2}\left(C_{2,0}^{\left(I\right)}\right)^{*}C_{4,0}^{\left(I\right)}\,,
\]

\[
\rho_{24}=\frac{1}{2}\left[\left(C_{1,0}^{\left(I\right)}\right)^{*}C_{3,0}^{\left(I\right)}+\left(C_{2,-1}^{\left(II\right)}\right)^{*}C_{4,-1}^{\left(II\right)}\right]\,.
\]

Thus the negativity $\mathcal{N}$ may be expressed as
\[
\mathcal{N}=2\sum_{i=1}^{2}\left|a_{i}\right|
\]
\[
a_{1}=\frac{1}{2}\left(\rho_{22}-\sqrt{\rho_{22}^{2}+4\left|\rho_{35}\right|^{2}}\right)\,,\qquad a_{2}=
\frac{1}{2}\left(\rho_{11}+\rho_{55}-\sqrt{(\rho_{11}-\rho_{55})^{2}+4\left|\rho_{24}\right|^{2}}\right)\,.
\]

\begin{figure}
\begin{centering}
\includegraphics[scale=0.9]{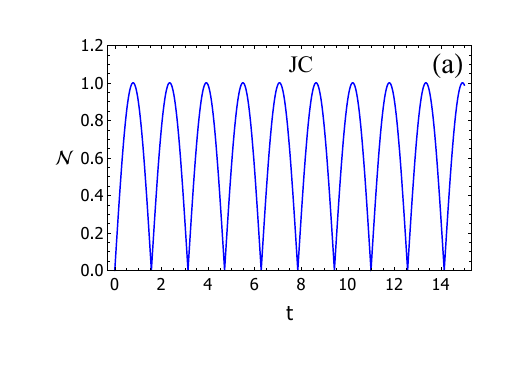}\includegraphics[scale=0.9]{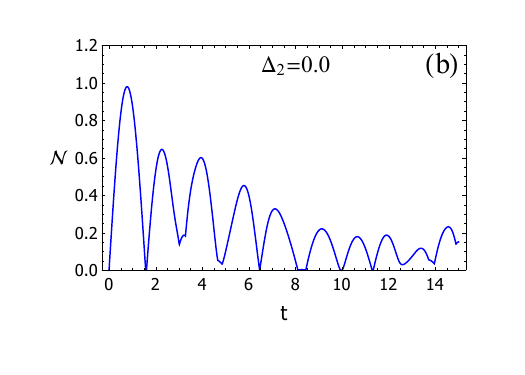}
\includegraphics[scale=0.9]{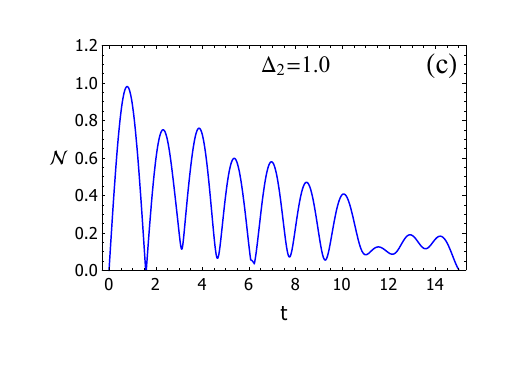}\includegraphics[scale=0.9]{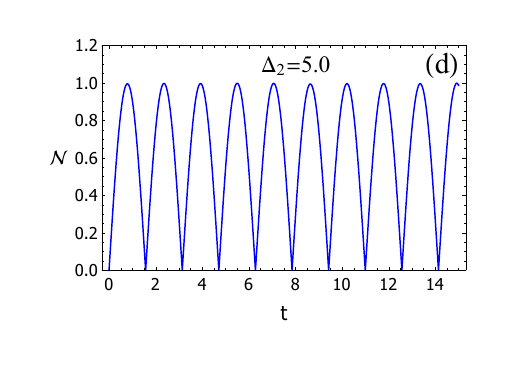}
\par\end{centering}

\caption{\label{fig:negativity} Negativity $\cal{N}$ used to quantify the entanglement between atom 1 and the field 
as a function of time. In (a) the field is not coupled to the environment, i.e.,
$\lambda_{2}=0.0$ (Jaynes Cummings model). For the field coupled to the environment, with $\lambda_{2}=0.2$
and (b) $\Delta_{2}=0.0$; (c) $\Delta_{2}=1.0$, and (d) $\Delta_{2}=5.0$.
In all cases $\lambda_{1}=1.0$, and the initial state of the system is a tripartite product state with
$\left|\psi_{a1}\right\rangle =\left|e_{1}\right\rangle $,
$\rho_{a2}=\frac{1}{2}\left(\left|g_{2}\right\rangle \left\langle g_{2}\right|+\left|e_{2}\right\rangle \left\langle e_{2}\right|\right)$
and $\left|\psi_{f}\right\rangle =\left|0\right\rangle $. }\label{fig3}
\end{figure}

We have inserted a factor of $2$ so that the maximum entanglement corresponds to $\mathcal{N} = 1$. We note that
although the negativity is somehow related to the linear entropy, they may not have the same behaviour, given that in this case
the reduced atom 1-field density operator is obtained from a mixed state. In figure (\ref{fig3}) we have plotted the negativity 
$\cal{N}$ as a function of time. For the initial conditions chosen here, the system has a simple evolution, i.e., a oscillatory 
behaviour as shown in figure (\ref{fig3}a), provided the system is isolated (decoupled environment). However, if atom 2 (environment) 
is resonantly coupled to the field, the entanglement is severely damped [see figure (\ref{fig3}b)], similarly to what happens 
to the linear entropy and atomic dipole squeezing. Again, entanglement may be restored if $\Delta_2$ is increased, as shown in 
figures (\ref{fig3}c) and (\ref{fig3}d).

\section{Time-averaged evolution of the linear entropy}

\begin{figure}
\begin{centering}
\includegraphics[scale=0.9]{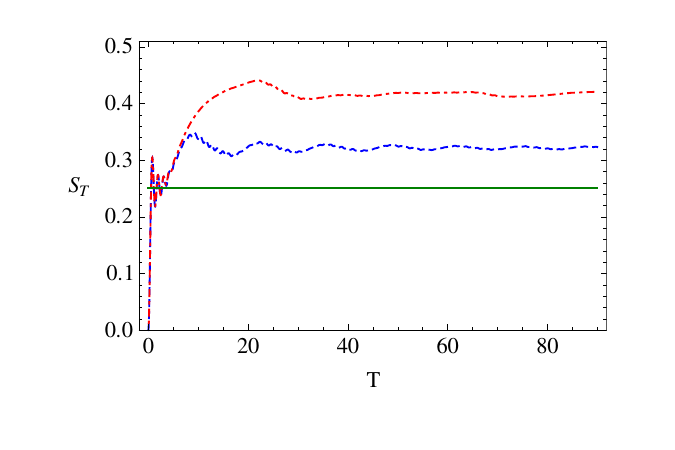}
\par\end{centering}
\caption{\label{fig:average}Time-averaged linear entropy ($S_T$) of atom 1, as a function of
time for a) $p=0.0$ dashed (blue) curve, and b) $p=0.5$ dot-dashed (red) curve. In both cases
$\lambda_{1}=1.0$, $\lambda_{2}=0.1$ and $\Delta_{1}=\Delta_{2}=0.0$. We also have
$\left|\psi_{a1}\right\rangle =\left|e_{1}\right\rangle $,
$\rho_{a2}=p\left|e_{2}\right\rangle \left\langle e_{2}\right|+(1-p)\left|g_{2}\right\rangle \left\langle g_{2}\right|$
and $\left|\psi_{f}\right\rangle =\left|1\right\rangle $. The continuous
green line is drawn for $S_T = 0.25$, corresponding to the uncoupled case, or $\lambda_{2}=0.0$}\label{fig4}
\end{figure}

Now we would like to estimate the influence of the degree of mixedness of the small environment in 
the decoherence process. A simple way of doing that is for instance by calculating the time average of the 
linear entropy, here denoted as $S_{T}$ and defined as
\begin{equation}
S_{T}\left(T\right)=\frac{1}{T}\int_{0}^{T}S\left(t\right)dt,
\end{equation}
where $S(t)$ the linear entropy in equation (\ref{eq:entropiatc}). We are not including the resulting 
expression for $S_{T}$ because it is a rather large one. 
We calculated the time average of the linear entropy of atom 1 in two distinct situations; for atom 2 
initially i) in a pure state ($p=0.0$) and ii) in a maximally mixed state ($p=0.5$).
Naturally, larger values of $S_{T}$ (steady state) are an indication of larger degradation of quantum coherence. 
In figure \ref{fig:average} we have plotted $S_{T}$ (defined above) as a function of time ($T$).
The dashed (blue) curve is for $p=0.0$ and the dot-dashed (red) curve for $p=0.5$. The influence of the degree of
mixedness of the small environment on the state purity of atom 1 is clear, as we note that there is a significant 
increase ($\sim 30\%$) in its linear entropy (for long times), if the small environment is initially in a maximally 
mixed state ($p=0.5$) rather than in a pure state ($p=0.0$). Note that the average value of the linear entropy in
the absence of an environment ($\lambda_2 = 0.0$) is $S_0 = 0.25$ (continuous green line in the plot).  
We would like to remark that a time averaging procedure may also be useful to discuss the effects 
of additional external fluctuations. Interestingly, solely by performing some kind of time-averaging it is possible 
to build a model for non-dissipative decoherence, as described in reference \cite{bonifacio00}. Also, a long-time
average of the density operator allows the investigation of the thermalization properties of systems having a small 
number of degrees of freedom, as shown in \cite{ueda17}. 

\section{Conclusions}

We have presented a study of the dynamics of a bipartite quantum system (atom 1$+$field) in which the field is weakly 
coupled ($\lambda_2 \approx \lambda_1/10$) to a small environment (atom 2). A single atom corresponds to the smallest 
possible environment considering the atomic beam model for a reservoir \cite{lambscully74}. 
Of course there is no relaxation in this simple model, contrarily to what happens in the case of a ``many atoms reservoir". 
Instead, due to the incommensurate frequencies characteristic of the model, we expect 
quasi-recurrences of the considered physical quantities at longer time-scales. For instance, as we have seen, the 
linear entropy of atom 1 undergoes an irreversible-like evolution. We have also investigated the evolution of quantities 
related to non-classical behaviour, such as the dipole squeezing of atom 1 and the negativity, which quantifies the 
atom 1-field quantum entanglement. We have found that both the atomic dipole squeezing and the atom 1-field 
entanglement are considerably degraded at shorter time scales (especially in the resonant case, $\Delta_2 = 0.0$).
In particular, we have verified that dipole squeezing, which occurs at relatively narrow time intervals, is 
completely suppressed. Of course due to the nature of the toy model here studied, there might occur quasi-revivals at 
longer times. However, even if a small amount of noise is added to the system, the joint action of the minimal 
environment and the extra noise could completely destroy the above mentioned quantum properties, as discussed in 
\cite{avb14}. 

Here we have studied a general case, by analytically solving the two-atom TCM for non-identical atoms, with 
different coupling constants, ($\lambda_1\neq\lambda_2$) and different detunings ($\Delta_1\neq\Delta_2$).
The solution of the two-atom TCM for non-identical atoms allowed us to assess the effect of the atom 2-field 
detuning on the dynamics of the system. Indeed there is a competition between the field-environment coupling and 
the field-environment detuning, but although we were able to determine explicitly the dependence of the quantum 
state of the system on the parameters ($\lambda_1,\lambda_2,\Delta_1,\Delta_2$), the lengthy expressions hindered 
a more detailed analysis. We have shown that it is possible to have some degree of control over the dynamics 
of the system and circumvent the destructive effect of the environment. By increasing the atom 2-field detuning 
$\Delta_2$, there will be an effective decoupling of the small environment, and we expect a restoration of the 
non-classical properties such as squeezing and entanglement. For instance, if $\Delta_2 = 1.0$ 
it is possible to recover the atomic dipole squeezing property of atom 1. Yet, larger values of detuning would 
be required to restore the atom-field entanglement, as it is clearly seen in figure (\ref{fig3}). 

We have also made an estimate of the influence of the degree of mixedness of the small environment on 
the evolution of the main system by calculating the time-average of the linear entropy for different effective 
temperatures of the environment.

\section*{Acknowledgements}

G.L.D. would like to thank CAPES (Coordena\c c\~ao de Aperfei\c coamento 
de Pessoal de N\'\i vel Superior) under grant 2011/899872, 
for financial support. This work was also supported by CNPq 
(Conselho Nacional para o Desenvolvimento Cient\'\i fico e Tecnol\'ogico) 
and FAPESP (Funda\c c\~ao de Amparo \`a Pesquisa do Estado de S\~ao Paulo), 
through the INCT-IQ (National Institute for Science and Technology of Quantum Information) 
under grant 2008/57856-6 and the CePOF (Optics and Photonics 
Research Center) under grant 2005/51689-2, Brazil.

\section*{Appendix. Solution of the two-atom Tavis-Cummings model}

Here we present the full analytical solution of the two-atom Tavis-Cummings model (coefficients $A_{kl}$).
The diagonal terms are 

\begin{eqnarray*}
A_{11}  =  \frac{1}{54}\left\{ \frac{e^{-\frac{1}{2}t\left(\sqrt{X}-\sqrt{Y_{1}}\right)}\left(18\left(\sqrt{3}+i\right)D_{1}J_{1}\left(-2ik_{1}+3\sqrt{X}-3\sqrt{Y_{1}}\right)-432D_{1}^{3}-4iJ_{1}^{3}+\left(2ik_{1}-3\sqrt{X}+3\sqrt{Y_{1}}\right){}^{3}\right)}{\sqrt{Y_{1}}\left(-4\sqrt{X}\sqrt{Y_{1}}+4X+Y_{1}-Y_{2}\right)}\right.\\
+\frac{e^{-\frac{1}{2}t\left(\sqrt{X}+\sqrt{Y_{1}}\right)}\left(18\left(\sqrt{3}+i\right)iD_{1}J_{1}\left(2k_{1}+3i\left(\sqrt{X}+\sqrt{Y_{1}}\right)\right)+432D_{1}^{3}+i\left(4J_{1}^{3}+\left(2k_{1}+3i\left(\sqrt{X}+\sqrt{Y_{1}}\right)\right){}^{3}\right)\right)}{\sqrt{Y_{1}}\left(4\sqrt{X}\sqrt{Y_{1}}+4X+Y_{1}-Y_{2}\right)}\\
  +\frac{e^{\frac{1}{2}t\left(\sqrt{X}-\sqrt{Y_{2}}\right)}\left(18\left(\sqrt{3}+i\right)D_{1}J_{1}\left(2ik_{1}+3\sqrt{X}-3\sqrt{Y_{2}}\right)+432D_{1}^{3}+4iJ_{1}^{3}+\left(-2ik_{1}-3\sqrt{X}+3\sqrt{Y_{2}}\right){}^{3}\right)}{\sqrt{Y_{2}}\left(-4\sqrt{X}\sqrt{Y_{2}}+4X-Y_{1}+Y_{2}\right)}\\
   \left.+\frac{e^{\frac{1}{2}t\left(\sqrt{X}+\sqrt{Y_{2}}\right)}\left(18\left(1-i\sqrt{3}\right)D_{1}J_{1}\left(2k_{1}-3i\left(\sqrt{X}+\sqrt{Y_{2}}\right)\right)-432D_{1}^{3}-i\left(4J_{1}^{3}+\left(2k_{1}-3i\left(\sqrt{X}+\sqrt{Y_{2}}\right)\right){}^{3}\right)\right)}{\sqrt{Y_{2}}\left(4\sqrt{X}\sqrt{Y_{2}}+4X-Y_{1}+Y_{2}\right)}\right\},
\end{eqnarray*}
\\

\begin{eqnarray*}
  A_{22}  =  \frac{1}{54}\left\{ \frac{e^{-\frac{1}{2}t\left(\sqrt{X}-\sqrt{Y_{1}}\right)}\left(18\left(\sqrt{3}+i\right)D_{2}J_{2}\left(-2ik_{2}+3\sqrt{X}-3\sqrt{Y_{1}}\right)-432D_{2}^{3}-4iJ_{2}^{3}+\left(2ik_{2}-3\sqrt{X}+3\sqrt{Y_{1}}\right){}^{3}\right)}{\sqrt{Y_{1}}\left(-4\sqrt{X}\sqrt{Y_{1}}+4X+Y_{1}-Y_{2}\right)}\right.\\
  +\frac{e^{-\frac{1}{2}t\left(\sqrt{X}+\sqrt{Y_{1}}\right)}\left(18\left(\sqrt{3}+i\right)iD_{2}J_{2}\left(2k_{2}+3i\left(\sqrt{X}+\sqrt{Y_{1}}\right)\right)+432D_{2}^{3}+i\left(4J_{2}^{3}+\left(2k_{2}+3i\left(\sqrt{X}+\sqrt{Y_{1}}\right)\right){}^{3}\right)\right)}{\sqrt{Y_{1}}\left(4\sqrt{X}\sqrt{Y_{1}}+4X+Y_{1}-Y_{2}\right)}\\
  +\frac{e^{\frac{1}{2}t\left(\sqrt{X}-\sqrt{Y_{2}}\right)}\left(18\left(\sqrt{3}+i\right)D_{2}J_{2}\left(2ik_{2}+3\sqrt{X}-3\sqrt{Y_{2}}\right)+432D_{2}^{3}+4iJ_{2}^{3}+\left(-2ik_{2}-3\sqrt{X}+3\sqrt{Y_{2}}\right){}^{3}\right)}{\sqrt{Y_{2}}\left(-4\sqrt{X}\sqrt{Y_{2}}+4X-Y_{1}+Y_{2}\right)}\\
  \left.+\frac{e^{\frac{1}{2}t\left(\sqrt{X}+\sqrt{Y_{2}}\right)}\left(18\left(1-i\sqrt{3}\right)D_{2}J_{2}\left(2k_{2}-3i\left(\sqrt{X}+\sqrt{Y_{2}}\right)\right)-432D_{2}^{3}-i\left(4J_{2}^{3}+\left(2k_{2}-3i\left(\sqrt{X}+\sqrt{Y_{2}}\right)\right){}^{3}\right)\right)}{\sqrt{Y_{2}}\left(4\sqrt{X}\sqrt{Y_{2}}+4X-Y_{1}+Y_{2}\right)}\right\},
\end{eqnarray*}
\\

\begin{eqnarray*}
  A_{33}  =\frac{1}{54}  \left\{ \frac{e^{-\frac{1}{2}t\left(\sqrt{X}-\sqrt{Y_{1}}\right)}\left(18\left(\sqrt{3}+i\right)D_{3}J_{3}\left(-2ik_{3}+3\sqrt{X}-3\sqrt{Y_{1}}\right)-432D_{3}^{3}-4iJ_{3}^{3}+\left(2ik_{3}-3\sqrt{X}+3\sqrt{Y_{1}}\right){}^{3}\right)}{\sqrt{Y_{1}}\left(-4\sqrt{X}\sqrt{Y_{1}}+4X+Y_{1}-Y_{2}\right)}\right.\\
  +\frac{e^{-\frac{1}{2}t\left(\sqrt{X}+\sqrt{Y_{1}}\right)}\left(18\left(\sqrt{3}+i\right)iD_{3}J_{3}\left(2k_{3}+3i\left(\sqrt{X}+\sqrt{Y_{1}}\right)\right)+432D_{3}^{3}+i\left(4J_{3}^{3}+\left(2k_{3}+3i\left(\sqrt{X}+\sqrt{Y_{1}}\right)\right){}^{3}\right)\right)}{\sqrt{Y_{1}}\left(4\sqrt{X}\sqrt{Y_{1}}+4X+Y_{1}-Y_{2}\right)}\\
  +\frac{e^{\frac{1}{2}t\left(\sqrt{X}-\sqrt{Y_{2}}\right)}\left(18\left(\sqrt{3}+i\right)D_{3}J_{3}\left(2ik_{3}+3\sqrt{X}-3\sqrt{Y_{2}}\right)+432D_{3}^{3}+4iJ_{3}^{3}+\left(-2ik_{3}-3\sqrt{X}+3\sqrt{Y_{2}}\right){}^{3}\right)}{\sqrt{Y_{2}}\left(-4\sqrt{X}\sqrt{Y_{2}}+4X-Y_{1}+Y_{2}\right)}\\
  \left.+\frac{e^{\frac{1}{2}t\left(\sqrt{X}+\sqrt{Y_{2}}\right)}\left(18\left(1-i\sqrt{3}\right)D_{3}J_{3}\left(2k_{3}-3i\left(\sqrt{X}+\sqrt{Y_{2}}\right)\right)-432D_{3}^{3}-i\left(4J_{3}^{3}+\left(2k_{3}-3i\left(\sqrt{X}+\sqrt{Y_{2}}\right)\right){}^{3}\right)\right)}{\sqrt{Y_{2}}\left(4\sqrt{X}\sqrt{Y_{2}}+4X-Y_{1}+Y_{2}\right)}\right\},
\end{eqnarray*}
\\

\begin{eqnarray*}
  A_{44}  =  \frac{1}{54}\left\{ \frac{e^{-\frac{1}{2}t\left(\sqrt{X}-\sqrt{Y_{1}}\right)}\left(18\left(\sqrt{3}+i\right)D_{4}J_{4}\left(-2ik_{4}+3\sqrt{X}-3\sqrt{Y_{1}}\right)-432D_{4}^{3}-4iJ_{4}^{3}+\left(2ik_{4}-3\sqrt{X}+3\sqrt{Y_{1}}\right){}^{3}\right)}{\sqrt{Y_{1}}\left(-4\sqrt{X}\sqrt{Y_{1}}+4X+Y_{1}-Y_{2}\right)}\right.\\
  +\frac{e^{-\frac{1}{2}t\left(\sqrt{X}+\sqrt{Y_{1}}\right)}\left(18\left(\sqrt{3}+i\right)iD_{4}J_{4}\left(2k_{4}+3i\left(\sqrt{X}+\sqrt{Y_{1}}\right)\right)+432D_{4}^{3}+i\left(4J_{4}^{3}+\left(2k_{4}+3i\left(\sqrt{X}+\sqrt{Y_{1}}\right)\right){}^{3}\right)\right)}{\sqrt{Y_{1}}\left(4\sqrt{X}\sqrt{Y_{1}}+4X+Y_{1}-Y_{2}\right)}\\
  +\frac{e^{\frac{1}{2}t\left(\sqrt{X}-\sqrt{Y_{2}}\right)}\left(18\left(\sqrt{3}+i\right)D_{4}J_{4}\left(2ik_{4}+3\sqrt{X}-3\sqrt{Y_{2}}\right)+432D_{4}^{3}+4iJ_{4}^{3}+\left(-2ik_{4}-3\sqrt{X}+3\sqrt{Y_{2}}\right){}^{3}\right)}{\sqrt{Y_{2}}\left(-4\sqrt{X}\sqrt{Y_{2}}+4X-Y_{1}+Y_{2}\right)}\\
  \left.+\frac{e^{\frac{1}{2}t\left(\sqrt{X}+\sqrt{Y_{2}}\right)}\left(18\left(1-i\sqrt{3}\right)D_{4}J_{4}\left(2k_{4}-3i\left(\sqrt{X}+\sqrt{Y_{2}}\right)\right)-432D_{4}^{3}-i\left(4J_{4}^{3}+\left(2k_{4}-3i\left(\sqrt{X}+\sqrt{Y_{2}}\right)\right){}^{3}\right)\right)}{\sqrt{Y_{2}}\left(4\sqrt{X}\sqrt{Y_{2}}+4X-Y_{1}+Y_{2}\right)}\right\},
\end{eqnarray*}

and the non-diagonal terms read 

\begin{eqnarray*}
  A_{14}=-4\lambda_{1}\lambda_{2}\sqrt{(n+1)(n+2)}\left\{ \frac{\left(\sqrt{Y_{1}}-\sqrt{X}\right)e^{-\frac{1}{2}t\left(\sqrt{X}-\sqrt{Y_{1}}\right)}}{\sqrt{Y_{1}}\left(-4\sqrt{X}\sqrt{Y_{1}}+4X+Y_{1}-Y_{2}\right)}+\frac{\left(\sqrt{X}+\sqrt{Y_{1}}\right)e^{-\frac{1}{2}t\left(\sqrt{X}+\sqrt{Y_{1}}\right)}}{\sqrt{Y_{1}}\left(4\sqrt{X}\sqrt{Y_{1}}+4X+Y_{1}-Y_{2}\right)}\right.\\
  \left.+\frac{\left(\sqrt{X}+\sqrt{Y_{2}}\right)e^{\frac{1}{2}t\left(\sqrt{X}+\sqrt{Y_{2}}\right)}}{\sqrt{Y_{2}}\left(4\sqrt{X}\sqrt{Y_{2}}+4X-Y_{1}+Y_{2}\right)}+\frac{\left(\sqrt{Y_{2}}-\sqrt{X}\right)e^{\frac{1}{2}t\left(\sqrt{X}-\sqrt{Y_{2}}\right)}}{\sqrt{Y_{2}}\left(-4\sqrt{X}\sqrt{Y_{2}}+4X-Y_{1}+Y_{2}\right)}\right\},
\end{eqnarray*}
\\

\begin{eqnarray*}
  A_{23}  =  -\frac{2e^{\frac{1}{2}t\left(\sqrt{X}+\sqrt{Y_{2}}\right)}\left(i\left(\Delta_{1}+\Delta_{2}\right)\lambda_{1}\lambda_{2}+(2n+3)\sqrt{X}+(2n+3)\sqrt{Y_{2}}\right)}{\sqrt{Y_{2}}\left(4\sqrt{X}\sqrt{Y_{2}}+4X-Y_{1}+Y_{2}\right)}\\
  +\frac{e^{-\frac{1}{2}t\left(\sqrt{X}+\sqrt{Y_{1}}\right)}\left(2i\left(\Delta_{1}+\Delta_{2}\right)\lambda_{1}\lambda_{2}-2(2n+3)\sqrt{X}-2(2n+3)\sqrt{Y_{1}}\right)}{\sqrt{Y_{1}}\left(4\sqrt{X}\sqrt{Y_{1}}+4X+Y_{1}-Y_{2}\right)}\\
  +\frac{e^{\frac{1}{2}t\left(\sqrt{X}-\sqrt{Y_{2}}\right)}\left(2i\left(\Delta_{1}+\Delta_{2}\right)\lambda_{1}\lambda_{2}+(4n+6)\sqrt{X}-2(2n+3)\sqrt{Y_{2}}\right)}{\sqrt{Y_{2}}\left(-4\sqrt{X}\sqrt{Y_{2}}+4X-Y_{1}+Y_{2}\right)}\\
  +\frac{e^{-\frac{1}{2}t\left(\sqrt{X}-\sqrt{Y_{1}}\right)}\left((4n+6)\sqrt{X}-2\left((2n+3)\sqrt{Y_{1}}+i\left(\Delta_{1}+\Delta_{2}\right)\lambda_{1}\lambda_{2}\right)\right)}{\sqrt{Y_{1}}\left(-4\sqrt{X}\sqrt{Y_{1}}+4X+Y_{1}-Y_{2}\right)},
\end{eqnarray*}
\\

\begin{eqnarray*}
  A_{13}  =  -i\lambda_{1}\sqrt{n+1}\left\{ -\frac{e^{\frac{1}{2}t\left(\sqrt{X}-\sqrt{Y_{2}}\right)}\left(G+\left(\sqrt{X}-\sqrt{Y_{2}}\right)\left(-2i\Delta_{2}+\sqrt{X}-\sqrt{Y_{2}}\right)\right)}{\sqrt{Y_{2}}\left(-4\sqrt{X}\sqrt{Y_{2}}+4X-Y_{1}+Y_{2}\right)}\right.\\
  +\frac{e^{\frac{1}{2}t\left(\sqrt{X}+\sqrt{Y_{2}}\right)}\left(G+\left(\sqrt{X}+\sqrt{Y_{2}}\right)\left(-2i\Delta_{2}+\sqrt{X}+\sqrt{Y_{2}}\right)\right)}{\sqrt{Y_{2}}\left(4\sqrt{X}\sqrt{Y_{2}}+4X-Y_{1}+Y_{2}\right)}\\
  +\frac{e^{-\frac{1}{2}t\left(\sqrt{X}-\sqrt{Y_{1}}\right)}\left(G+\left(\sqrt{X}-\sqrt{Y_{1}}\right)\left(2i\Delta_{2}+\sqrt{X}-\sqrt{Y_{1}}\right)\right)}{\sqrt{Y_{1}}\left(-4\sqrt{X}\sqrt{Y_{1}}+4X+Y_{1}-Y_{2}\right)}\\
  \left.+\frac{e^{-\frac{1}{2}t\left(\sqrt{X}+\sqrt{Y_{1}}\right)}\left(-G-\left(\sqrt{X}+\sqrt{Y_{1}}\right)\left(2i\Delta_{2}+\sqrt{X}+\sqrt{Y_{1}}\right)\right)}{\sqrt{Y_{1}}\left(4\sqrt{X}\sqrt{Y_{1}}+4X+Y_{1}-Y_{2}\right)}\right\},
\end{eqnarray*}
\\

\begin{eqnarray*}
  A_{12}  =  -i\sqrt{n+1}\lambda_{2}\left\{ \frac{e^{\frac{1}{2}t\left(\sqrt{X}-\sqrt{Y_{2}}\right)}\left(G-\left(\sqrt{X}-\sqrt{Y_{2}}\right)\left(-2i\Delta_{1}+\sqrt{X}-\sqrt{Y_{2}}\right)\right)}{\sqrt{Y_{2}}\left(-4\sqrt{X}\sqrt{Y_{2}}+4X-Y_{1}+Y_{2}\right)}\right.\\
  +\frac{e^{\frac{1}{2}t\left(\sqrt{X}+\sqrt{Y_{2}}\right)}\left(-G+\left(\sqrt{X}+\sqrt{Y_{2}}\right)\left(-2i\Delta_{1}+\sqrt{X}+\sqrt{Y_{2}}\right)\right)}{\sqrt{Y_{2}}\left(4\sqrt{X}\sqrt{Y_{2}}+4X-Y_{1}+Y_{2}\right)}\\
  +\frac{e^{-\frac{1}{2}t\left(\sqrt{X}-\sqrt{Y_{1}}\right)}\left(-G+\left(\sqrt{X}-\sqrt{Y_{1}}\right)\left(2i\Delta_{1}+\sqrt{X}-\sqrt{Y_{1}}\right)\right)}{\sqrt{Y_{1}}\left(-4\sqrt{X}\sqrt{Y_{1}}+4X+Y_{1}-Y_{2}\right)}\\
  \left.+\frac{e^{-\frac{1}{2}t\left(\sqrt{X}+\sqrt{Y_{1}}\right)}\left(G-\left(\sqrt{X}+\sqrt{Y_{1}}\right)\left(2i\Delta_{1}+\sqrt{X}+\sqrt{Y_{1}}\right)\right)}{\sqrt{Y_{1}}\left(4\sqrt{X}\sqrt{Y_{1}}+4X+Y_{1}-Y_{2}\right)}\right\},
\end{eqnarray*}
\\

\begin{eqnarray*}
  A_{24}  =  -i\sqrt{n+2}\lambda_{1}\left\{ \frac{e^{-\frac{1}{2}t\left(\sqrt{X}-\sqrt{Y_{1}}\right)}\left(H+\left(\sqrt{X}-\sqrt{Y_{1}}\right)\left(-2i\Delta_{2}+\sqrt{X}-\sqrt{Y_{1}}\right)\right)}{\sqrt{Y_{1}}\left(-4\sqrt{X}\sqrt{Y_{1}}+4X+Y_{1}-Y_{2}\right)}\right.\\
  +\frac{e^{-\frac{1}{2}t\left(\sqrt{X}+\sqrt{Y_{1}}\right)}\left(-H-\left(\sqrt{X}+\sqrt{Y_{1}}\right)\left(-2i\Delta_{2}+\sqrt{X}+\sqrt{Y_{1}}\right)\right)}{\sqrt{Y_{1}}\left(4\sqrt{X}\sqrt{Y_{1}}+4X+Y_{1}-Y_{2}\right)}\\
  +\frac{e^{\frac{1}{2}t\left(\sqrt{X}+\sqrt{Y_{2}}\right)}\left(H+\left(\sqrt{X}+\sqrt{Y_{2}}\right)\left(2i\Delta_{2}+\sqrt{X}+\sqrt{Y_{2}}\right)\right)}{\sqrt{Y_{2}}\left(4\sqrt{X}\sqrt{Y_{2}}+4X-Y_{1}+Y_{2}\right)}\\
  \left.-\frac{e^{\frac{1}{2}t\left(\sqrt{X}-\sqrt{Y_{2}}\right)}\left(H+\left(\sqrt{X}-\sqrt{Y_{2}}\right)\left(2i\Delta_{2}+\sqrt{X}-\sqrt{Y_{2}}\right)\right)}{\sqrt{Y_{2}}\left(-4\sqrt{X}\sqrt{Y_{2}}+4X-Y_{1}+Y_{2}\right)}\right\},
\end{eqnarray*}
\\

\begin{eqnarray*}
  A_{34}  =  -i\sqrt{n+2}\lambda_{2}\left\{ \frac{e^{-\frac{1}{2}t\left(\sqrt{X}-\sqrt{Y_{1}}\right)}\left(-H+\left(\sqrt{X}-\sqrt{Y_{1}}\right)\left(-2i\Delta_{1}+\sqrt{X}-\sqrt{Y_{1}}\right)\right)}{\sqrt{Y_{1}}\left(-4\sqrt{X}\sqrt{Y_{1}}+4X+Y_{1}-Y_{2}\right)}\right.\\
  +\frac{e^{-\frac{1}{2}t\left(\sqrt{X}+\sqrt{Y_{1}}\right)}\left(H-\left(\sqrt{X}+\sqrt{Y_{1}}\right)\left(-2i\Delta_{1}+\sqrt{X}+\sqrt{Y_{1}}\right)\right)}{\sqrt{Y_{1}}\left(4\sqrt{X}\sqrt{Y_{1}}+4X+Y_{1}-Y_{2}\right)}\\
  +\frac{e^{\frac{1}{2}t\left(\sqrt{X}-\sqrt{Y_{2}}\right)}\left(H-\left(\sqrt{X}-\sqrt{Y_{2}}\right)\left(2i\Delta_{1}+\sqrt{X}-\sqrt{Y_{2}}\right)\right)}{\sqrt{Y_{2}}\left(-4\sqrt{X}\sqrt{Y_{2}}+4X-Y_{1}+Y_{2}\right)}\\
  \left.+\frac{e^{\frac{1}{2}t\left(\sqrt{X}+\sqrt{Y_{2}}\right)}\left(-H+\left(\sqrt{X}+\sqrt{Y_{2}}\right)\left(2i\Delta_{1}+\sqrt{X}+\sqrt{Y_{2}}\right)\right)}{\sqrt{Y_{2}}\left(4\sqrt{X}\sqrt{Y_{2}}+4X-Y_{1}+Y_{2}\right)}\right\}.
\end{eqnarray*}

In the expressions above, the coefficients $X$ and $Y_{1,\,2}$ are given by

\[
X=-\frac{2A}{3}+\frac{2^{1/3}\left(A^{2}+12F\right)}{3\beta}+\frac{\Gamma}{3\times2^{1/3}}
\]

\[
Y_{1,\,2}=-2A-X\pm\frac{2iB}{\sqrt{X}},
\]

where 

\[
\Gamma=\left[2A^{3}-27B^{2}-72AF+\sqrt{-4\left(A^{2}+12F\right)^{3}+\left(2A^{3}-27B^{2}-72AF\right)^{2}}\right]^{1/3},
\]

and

\[
A=\frac{1}{2}\left(\Delta_{1}^{2}+\Delta_{2}^{2}+2(2n+3)\left(\lambda_{1}^{2}+\lambda_{2}^{2}\right)\right),
\]

\[
B=\left(\Delta_{1}\lambda_{2}^{2}+\Delta_{2}\lambda_{1}^{2}\right),
\]

\begin{eqnarray*}
F & = & \frac{1}{16}\left\{ \left(\Delta_{1}^{2}-\Delta_{2}^{2}\right)\left[\left(\Delta_{1}^{2}-\Delta_{2}^{2}\right)^{2}+4\left(3+2n\right)\left(\lambda_{1}^{2}-\lambda_{2}^{2}\right)\right]+16\left(n^{2}+3n+2\right)\left(\lambda_{1}^{2}-\lambda_{2}^{2}\right)^{2}\right\} .\\
\end{eqnarray*}

We have also that

\[
G=\Delta_{1}^{2}-\Delta_{2}^{2}+4\left(n+2\right)\left(\lambda_{1}^{2}-\lambda_{2}^{2}\right),
\]

\[
H=\Delta_{1}^{2}-\Delta_{2}^{2}+4\left(n+1\right)\left(\lambda_{1}^{2}-\lambda_{2}^{2}\right),
\]

\[
D_{i}=\frac{\left(-1\right)^{5/6}\left(3a_{i}+k_{i}^{2}\right)}{3J_{i}}\;,\quad i=1,\,2,\,3,\,4 ,
\]

and

\[
  J_{i}=\left(-27b_{i}+9a_{i}k_{i}+2k_{i}^{3}-3i\sqrt{3}\sqrt{4a_{i}^{3}-27b_{i}^{2}
+18a_{i}b_{i}k_{i}+a_{i}^{2}k_{i}^{2}+4b_{i}k_{i}^{3}}\right)^{1/3}\;,\quad i=1,\,2,\,3,\,4 ,
\]
being

\[
a_{1}=a_4=\frac{\left[\left(\Delta_{1}-\Delta_{2}\right)^{2}+4\left(n+2\right)\left(\lambda_{1}^{2}
+\lambda_{2}^{2}\right)\right]}{4},
\]

\[
a_{2}=a_3=\frac{\left[\left(\Delta_{1}+\Delta_{2}\right)^{2}+4\left(\left(n+1\right)\lambda_{1}^{2}
+\left(n+2\right)\lambda_{2}^{2}\right)\right]}{4},
\]

\[
b_{1} = -b_4 = -\frac{\left(\Delta_{1}-\Delta_{2}\right)}{8}\left[\Delta_{1}^{2}-\Delta_{2}^{2}
+4\left(n+2\right)\left(\lambda_{1}^{2}-\lambda_{2}^{2}\right)\right],
\]

\[
b_{2} = -b_3 = -\frac{\left(\Delta_{1}+\Delta_{2}\right)}{8}\left[\Delta_{1}^{2}-\Delta_{2}^{2}
+4\left(\left(n+1\right)\lambda_{1}^{2}-\left(n+2\right)\lambda_{2}^{2}\right)\right],
\]

\[
k_{1}=-k_4=-\frac{\left(\Delta_{1}+\Delta_{2}\right)}{2},
\]

and

\[
k_{2}=-k_3=-\frac{\left(\Delta_{1}-\Delta_{2}\right)}{2}.
\]

\newpage

\end{document}